\input harvmac
\let\includefigures=\iftrue
\newfam\black
\includefigures
\input epsf
\def\figin{\epsfcheck\figin}\def\figins{\epsfcheck\figins}
\def\epsfcheck{\ifx\epsfbox\UnDeFiNeD
\message{(NO epsf.tex, FIGURES WILL BE IGNORED)}
\gdef\figin##1{\vskip2in}\gdef\figins##1{\hskip.5in}
\else\message{(FIGURES WILL BE INCLUDED)}%
\gdef\figin##1{##1}\gdef\figins##1{##1}\fi}
\def\DefWarn#1{}
\def\figinsert{\goodbreak\midinsert}
\def\ifig#1#2#3{\DefWarn#1\xdef#1{fig.~\the\figno}
\writedef{#1\leftbracket fig.\noexpand~\the\figno}%
\figinsert\figin{\centerline{#3}}\medskip\centerline{\vbox{\baselineskip12pt
\advance\hsize by -1truein\noindent\footnotefont{\bf Fig.~\the\figno:}
#2}}
\bigskip\endinsert\global\advance\figno by1}
\else
\def\ifig#1#2#3{\xdef#1{fig.~\the\figno}
\writedef{#1\leftbracket fig.\noexpand~\the\figno}%
#2}}
\global\advance\figno by1}
\fi

\def\quad{{\ \ }}
\def\ZZ{{\bf Z}}
\def\hrho{{\hat \rho}}
\def\eps{\varepsilon}
\def\sign{{\rm sign}}
\def\ra{\rightarrow}
\def\frac#1#2{{{#1}\over {#2}}}
\def\Tr{{\rm Tr}}

\def\la{\langle}
\def\ra{\rangle}

\def\trho{{\tilde\rho}}
\def\tl{{\tilde l}}
\def\tT{{\tilde T}}
\def\tmu{{\tilde \mu}}
\lref\McGreevyKB{
J.~McGreevy and H.~Verlinde,
``Strings from tachyons: The c = 1 matrix reloated,''
arXiv:hep-th/0304224.
}
\lref\MorrisCQ{
T.~R.~Morris,
``Checkered Surfaces And Complex Matrices,''
Nucl.\ Phys.\ B {\bf 356}, 703 (1991).
}
\lref\AndersonNW{
A.~Anderson, R.~C.~Myers and V.~Periwal,
``Complex Random Surfaces,''
Phys.\ Lett.\ B {\bf 254}, 89 (1991).
}
\lref\dtrone{
S.~R.~Das, A.~Dhar, A.~M.~Sengupta and S.~R.~Wadia,
``New Critical Behavior In D = 0 Large N Matrix Models,''
Mod.\ Phys.\ Lett.\ A {\bf 5}, 1041 (1990).
}
\lref\dtrtwo{
S.~S.~Gubser and I.~R.~Klebanov,
``A Modified c = 1 matrix model with new critical behavior,''
Phys.\ Lett.\ B {\bf 340}, 35 (1994)
[arXiv:hep-th/9407014].
}
\lref\AharonyPA{
O.~Aharony, M.~Berkooz and E.~Silverstein,
``Multiple-trace operators and non-local string theories,''
JHEP {\bf 0108}, 006 (2001)
[arXiv:hep-th/0105309].
}
\lref\CMintegrone{
A.~P.~Polychronakos,
``Exchange Operator Formalism For Integrable Systems Of Particles,''
Phys.\ Rev.\ Lett.\  {\bf 69}, 703 (1992)
[arXiv:hep-th/9202057].
}
\lref\CMintegrtwo{
L.~Brink, T.~H.~Hansson and M.~A.~Vasiliev,
``Explicit solution to the N body Calogero problem,''
Phys.\ Lett.\ B {\bf 286}, 109 (1992)
[arXiv:hep-th/9206049].
}
\lref\CMBCone{
T.~Yamamoto, ``Multicomponent Calogero model of B(N) type confined in harmonic potential,''
arXiv:cond-mat/9508012.
}
\lref\CMBCtwo{
P.~K.~Ghosh, A.~Khare and M.~Sivakumar,
``Supersymmetry, Shape Invariance and Solvability of $A_{N-1}$ and $BC_{N}$ Calogero-Sutherland Model,''
Phys.\ Rev.\ A {\bf 58}, 821 (1998)
[arXiv:cond-mat/9710206].
}
\lref\genstatone{
F.~D.~Haldane, ``'Fractional Statistics' In Arbitrary Dimensions: A Generalization Of The Pauli Principle,''
Phys.\ Rev.\ Lett.\  {\bf 67}, 937 (1991).
}
\lref\genstattwo{
Y.~S.~Wu,
``Statistical distribution for generalized ideal gas of fractional statistics particles,''
Phys.\ Rev.\ Lett.\  {\bf 73}, 922 (1994).
}
\lref\AhnEV{
C.~Ahn, C.~Rim and M.~Stanishkov,
``Exact one-point function of N = 1 super-Liouville theory with boundary,''
Nucl.\ Phys.\ B {\bf 636}, 497 (2002)
[arXiv:hep-th/0202043].
}
\lref\FukudaBV{
T.~Fukuda and K.~Hosomichi,
``Super Liouville theory with boundary,''
Nucl.\ Phys.\ B {\bf 635}, 215 (2002)
[arXiv:hep-th/0202032].
}
\lref\Mehta{
M.~L.~Mehta,
``Random Matrices,''
Academic Press, 1991.
}
\lref\DiFrancescoUD{
P.~Di Francesco and D.~Kutasov,
``World sheet and space-time physics in two-dimensional (Super)string theory,''
Nucl.\ Phys.\ B {\bf 375}, 119 (1992)
[arXiv:hep-th/9109005].
}
\lref\GuptaFU{
A.~Gupta, S.~P.~Trivedi and M.~B.~Wise,
``Random Surfaces In Conformal Gauge,''
Nucl.\ Phys.\ B {\bf 340}, 475 (1990).
}
\lref\JevickiMB{
A.~Jevicki and B.~Sakita,
``The Quantum Collective Field Method And Its Application To The Planar  Limit,''
Nucl.\ Phys.\ B {\bf 165}, 511 (1980).
}
\lref\MooreSeiberg{
G.~W.~Moore and N.~Seiberg,
``From loops to fields in 2-D quantum gravity,''
Int.\ J.\ Mod.\ Phys.\ A {\bf 7}, 2601 (1992).
}
\lref\BergmanKM{
O.~Bergman and M.~R.~Gaberdiel,
``Dualities of type 0 strings,''
JHEP {\bf 9907}, 022 (1999)
[arXiv:hep-th/9906055].
}
\lref\DudasWD{
E.~Dudas, J.~Mourad and A.~Sagnotti,
``Charged and uncharged D-branes in various string theories,''
Nucl.\ Phys.\ B {\bf 620}, 109 (2002)
[arXiv:hep-th/0107081].
}

\lref\PolchinskiTU{
J.~Polchinski and Y.~Cai,
``Consistency Of Open Superstring Theories,''
Nucl.\ Phys.\ B {\bf 296}, 91 (1988).
}
\lref\CallanPX{
C.~G.~Callan, C.~Lovelace, C.~R.~Nappi and S.~A.~Yost,
``Adding Holes And Crosscaps To The Superstring,''
Nucl.\ Phys.\ B {\bf 293}, 83 (1987).
}

\lref\KlebanovKM{
I.~R.~Klebanov, J.~Maldacena and N.~Seiberg,
``D-brane decay in two-dimensional string theory,''
JHEP {\bf 0307}, 045 (2003)
[arXiv:hep-th/0305159].
}
\lref\McGreevyEP{
J.~McGreevy, J.~Teschner and H.~Verlinde,
``Classical and quantum D-branes in 2D string theory,''
arXiv:hep-th/0305194.
}
\lref\MyersYD{
R.~C.~Myers and V.~Periwal,
``Exactly Solvable Oriented Selfdual Strings,''
Phys.\ Rev.\ Lett.\  {\bf 64}, 3111 (1990).
}
\lref\SenMG{
A.~Sen,
``Non-BPS states and branes in string theory,''
arXiv:hep-th/9904207.
}
\lref\DornXN{
H.~Dorn and H.~J.~Otto,
``Two and three point functions in Liouville theory,''
Nucl.\ Phys.\ B {\bf 429}, 375 (1994)
[arXiv:hep-th/9403141].
}
\lref\MooreSF{
G.~W.~Moore,
``Double scaled field theory at c = 1,''
Nucl.\ Phys.\ B {\bf 368}, 557 (1992).
}
\lref\DornAT{
H.~Dorn and H.~J.~Otto,
``On correlation functions for noncritical strings with c <= 1 d >= 1,''
Phys.\ Lett.\ B {\bf 291}, 39 (1992)
[arXiv:hep-th/9206053].
}
\lref\AndricJK{
I.~Andric, A.~Jevicki and H.~Levine,
``On The Large N Limit In Symplectic Matrix Models,''
Nucl.\ Phys.\ B {\bf 215}, 307 (1983).
}

\lref\FischlerTB{
W.~Fischler and L.~Susskind,
``Dilaton Tadpoles, String Condensates And Scale Invariance. 2,''
Phys.\ Lett.\ B {\bf 173}, 262 (1986).
}
\lref\FischlerCI{
W.~Fischler and L.~Susskind,
``Dilaton Tadpoles, String Condensates And Scale Invariance,''
Phys.\ Lett.\ B {\bf 171}, 383 (1986).
}
\lref\GrossUB{
D.~J.~Gross and I.~R.~Klebanov,
``One-Dimensional String Theory On A Circle,''
Nucl.\ Phys.\ B {\bf 344}, 475 (1990).
}

\lref\ZamolodchikovAH{
A.~B.~Zamolodchikov and A.~B.~Zamolodchikov,
``Liouville field theory on a pseudosphere,''
arXiv:hep-th/0101152.
}

\lref\ZamolodchikovAA{
A.~B.~Zamolodchikov and A.~B.~Zamolodchikov,
``Structure constants and conformal bootstrap in Liouville field theory,''
Nucl.\ Phys.\ B {\bf 477}, 577 (1996)
[arXiv:hep-th/9506136].
}
\lref\SusskindKW{
L.~Susskind,
``The anthropic landscape of string theory,''
arXiv:hep-th/0302219.
}
\lref\Brezinetal{
E.~Brezin, C.~Itzykson, G.~Parisi and J.~B.~Zuber,
``Planar Diagrams,''
Commun.\ Math.\ Phys.\  {\bf 59}, 35 (1978).
}
\lref\PolychronakosSX{
A.~P.~Polychronakos,
arXiv:hep-th/9902157.
}
\lref\HikidaBT{
Y.~Hikida,
JHEP {\bf 0305}, 002 (2003)
[arXiv:hep-th/0210305].
}
\lref\DouglasUM{
M.~R.~Douglas,
``The statistics of string / M theory vacua,''
JHEP {\bf 0305}, 046 (2003)
[arXiv:hep-th/0303194].
}
\lref\TeschnerRV{
J.~Teschner,
``Liouville theory revisited,''
Class.\ Quant.\ Grav.\  {\bf 18}, R153 (2001)
[arXiv:hep-th/0104158].
}
\lref\NakayamaEP{
Y.~Nakayama,
``Tadpole cancellation in unoriented Liouville theory,''
arXiv:hep-th/0309063.
}

\lref\SenMH{
A.~Sen,
``Descent relations among bosonic D-branes,''
Int.\ J.\ Mod.\ Phys.\ A {\bf 14}, 4061 (1999)
[arXiv:hep-th/9902105].
}
\lref\TakayanagiSM{
T.~Takayanagi and N.~Toumbas,
``A matrix model dual of type 0B string theory in two dimensions,''
JHEP {\bf 0307}, 064 (2003)
[arXiv:hep-th/0307083].
}
\lref\DouglasUP{
M.~R.~Douglas, I.~R.~Klebanov, D.~Kutasov, J.~Maldacena, E.~Martinec and N.~Seiberg,
``A new hat for the c = 1 matrix model,''
arXiv:hep-th/0307195.
}
\lref\BershadskyXB{
M.~Bershadsky and I.~R.~Klebanov,
``Genus One Path Integral In Two-Dimensional Quantum Gravity,''
Phys.\ Rev.\ Lett.\  {\bf 65}, 3088 (1990).
}

\lref\KazakovCH{
V.~A.~Kazakov and A.~A.~Migdal,
``Recent Progress In The Theory Of Noncritical Strings,''
Nucl.\ Phys.\ B {\bf 311}, 171 (1988).
}
\lref\GinspargIS{
P.~Ginsparg and G.~W.~Moore,
``Lectures On 2-D Gravity And 2-D String Theory,''
arXiv:hep-th/9304011.
}
\lref\KlebanovQA{
I.~R.~Klebanov,
``String theory in two-dimensions,''
arXiv:hep-th/9108019.
}
\lref\PolchinskiMB{
J.~Polchinski,
``What is string theory?,''
arXiv:hep-th/9411028.
}
\lref\JevickiQN{
A.~Jevicki,
``Development in 2-d string theory,''
arXiv:hep-th/9309115.
}

\Title{\vbox{\baselineskip12pt\hbox{CALT-68-2458}\hbox{hep-th/0310195}}}
{\vbox{\centerline{
Two-Dimensional Unoriented Strings} \vskip2pt\centerline
{ And Matrix Models}}}

\centerline{Jaume Gomis and Anton Kapustin}
\bigskip\centerline{\it California Institute of Technology 452-48,
Pasadena, CA 91125}
\vskip .3in

\centerline{Abstract}

We investigate unoriented strings and superstrings in two dimensions and their dual matrix quantum mechanics. 
Most of the models we study have a tachyon tadpole coming from the $RP^2$ worldsheet which
needs to be cancelled by a renormalization of the worldsheet theory. We find evidence that the dual matrix models
describe the renormalized theory. The singlet sector of the matrix models 
is integrable and can be formulated in terms of fermions moving in an external potential and interacting via the
Calogero-Moser potential. 
We show that in the double-scaling limit the latter system exhibits particle-hole duality and interpret it
in terms of the dual string theory. We also show
that oriented string theories in two dimensions can be continuously deformed into unoriented ones
by turning on non-local interactions on the worldsheet.
We find two unoriented 
superstring models for which only oriented worldsheets contribute to
the S-matrix. A simple explanation for this is found in the dual matrix
model.

\Date{10/2003}
\newsec{Introduction}

The original connection between $D=2$  string theory and 
matrix quantum mechanics stems from the fact
that the  Feynman diagrams of the matrix model 
generate  in the double 
scaling limit the string
worldsheet expansion (see \KlebanovQA\GinspargIS\JevickiQN\PolchinskiMB\
for reviews and references). Recently, a more modern perspective  on this
connection -- now called duality -- has
emerged \McGreevyKB\KlebanovKM\McGreevyEP,  which relies on the
physics of tachyon 
condensation. In a nutshell, the matrix quantum mechanics is
identified with the worldvolume theory on unstable D0-branes in $D=2$
string theory. The connection then follows from the expectation that once
the branes disappear via tachyon condensation \SenMH\SenMG , one is
left with the 
original string theory but in the absence of branes and that in some
not fully understood sense the D0-branes can describe the closed
string background they were originally immersed in. This new
perspective has paved the way for supersymmetric generalizations
\TakayanagiSM\DouglasUP\  involving
$\hat{c}$=1 string theory\foot{A derivation of this duality using
a fishnet diagram expansion is still missing. It would be interesting to see how worldsheet
supersymmetry emerges from this point of view.}, resulting in
nonperturbatively consistent string backgrounds. For  related work see
\nref\MartinecKA{
E.~J.~Martinec,
``The annular report on non-critical string theory,''
arXiv:hep-th/0305148.
}
\nref\AlexandrovNN{
S.~Y.~Alexandrov, V.~A.~Kazakov and D.~Kutasov,
``Non-perturbative effects in matrix models and D-branes,''
JHEP {\bf 0309}, 057 (2003)
[arXiv:hep-th/0306177].
}
\nref\GaiottoYF{
D.~Gaiotto, N.~Itzhaki and L.~Rastelli,
``On the BCFT description of holes in the c = 1 matrix model,''
arXiv:hep-th/0307221.
}
\nref\GutperleIJ{
M.~Gutperle and P.~Kraus,
``D-brane dynamics in the c = 1 matrix model,''
arXiv:hep-th/0308047.
}
\nref\SenIV{
A.~Sen,
``Open-closed duality: Lessons from matrix model,''
arXiv:hep-th/0308068.
}
\nref\McGreevyDN{
J.~McGreevy, S.~Murthy and H.~Verlinde,
``Two-dimensional superstrings and the supersymmetric matrix model,''
arXiv:hep-th/0308105.
}
\nref\KapustinHI{
A.~Kapustin,
``Noncritical superstrings in a Ramond-Ramond background,''
arXiv:hep-th/0308119.
}
\nref\TeschnerQK{
J.~Teschner,
``On boundary perturbations in Liouville theory and brane dynamics in noncritical string theories,''
arXiv:hep-th/0308140.
}
\nref\GiveonWN{
A.~Giveon, A.~Konechny, A.~Pakman and A.~Sever,
``Type 0 strings in a 2-d black hole,''
arXiv:hep-th/0309056.
}
\lref\ShenkerUF{
S.~H.~Shenker,
``The Strength Of Nonperturbative Effects In String Theory,''
RU-90-47
{\it Presented at the Cargese Workshop on Random Surfaces, Quantum Gravity and Strings, Cargese, France, May 28 - Jun 1, 1990}
}
\nref\KarczmarekPV{
J.~L.~Karczmarek and A.~Strominger,
``Matrix cosmology,''
arXiv:hep-th/0309138.
}
\nref\DeWolfeQF{
O.~DeWolfe, R.~Roiban, M.~Spradlin, A.~Volovich and J.~Walcher,
``On the S-matrix of type 0 string theory,''
arXiv:hep-th/0309148.
}
\nref\KlebanovWG{
I.~R.~Klebanov, J.~Maldacena and N.~Seiberg,
``Unitary and complex matrix models as 1-d type 0 strings,''
arXiv:hep-th/0309168.
}
\nref\DasguptaKK{
S.~Dasgupta and T.~Dasgupta,
``Renormalization group approach to c = 1 matrix model on a circle and D-brane decay,''
arXiv:hep-th/0310106.
}
\nref\AlexandrovUN{
S.~Alexandrov,
``(m,n) ZZ branes and the c = 1 matrix model,''
arXiv:hep-th/0310135.
}
\hskip-45pt \MartinecKA-\AlexandrovUN. 

Studying these low dimensional models of string theory may seem a bit
academic. Nevertheless, their study in the early 90's
resulted in general lessons
applicable to more interesting (realistic) string vacua, most notably
Shenker' s estimates \ShenkerUF\  on the contribution of
non-perturbative effects. In this paper we consider various
two-dimensional models and extract two general lessons from the dual
matrix models, 
which we expect to apply to other string vacua, but that are hard to
establish with the present technology. The first general lesson is that models
with  NS-NS tadpoles can be consistently renormalized to yield
calculable, well defined results. Four dimensional, non-supersymmetric
models of particle physics constructed from branes generically have
NS-NS tadpoles and it is therefore an important issue to address the
treatment of tadpole divergences in
string theory. For two-dimensional string theory, the matrix model
duals give an
unambiguous procedure for how to get physically
sensible results, providing a hint that the problem with tadpole
divergences can be dealt with in higher dimensions. The other general
lesson is the high degree of 
connectivity of string vacua, by which vacua which look very different
from the worldsheet point of view can be continuously deformed into
each other. In the
two-dimensional context considered in this paper, 
we provide an example of a non-local
worldsheet interaction of the type studied in \AharonyPA\ which
interpolates between a model of oriented strings and a model with an
orientifold! Taking these deformations into account is important when 
discussing the landscape of string theory
vacua \SusskindKW\DouglasUM.

In this paper we consider $c=1$ and ${\hat c}=1$ string theory models of
unoriented strings, find the corresponding matrix model duals, and test
the duality by performing some perturbative computations. Most of the
models have a tachyon tadpole on the $RP^2$ worldsheet, which result in
divergences in the $RP^2$ contribution to the 
S-matrix and in the partition function. 
 Nevertheless, we find
evidence that the dual matrix models, which can be
reduced to a system of fermions interacting via a Calogero-Moser potential, actually
describe string theory 
propagation in a shifted background, where the string divergences
have been 
cancelled via the Fischler-Susskind \FischlerCI\FischlerTB\ mechanism.
The matrix model gives an unambiguous regularization of the string
divergences and appears to know about the final, finite, 
stable closed string vacuum of the model\foot{Even in the oriented
$c=1$ string model there are divergences due to a tachyon tadpole on
the $T^2$ worldsheet. Therefore, $T^2$ corrections to the S-matrix and
higher loop amplitudes are divergent. It is important to renormalize
these divergences before comparing with the matrix model amplitudes,
which are manifestly finite. It would be interesting to carry this out.}.

We  find that the matrix models for {\it unoriented} strings
can be reached by a
particular double-trace 
deformation of the familiar Hermitian matrix model, which describes
{\it oriented} strings in two dimensions. Using the dictionary between
oriented string theory and the Hermitian matrix model, it is possible to write down
a non-local worldsheet deformation \dtrone\dtrtwo\AharonyPA\
responsible for connecting the
vacuum with orientable strings to the vacuum with the
orientifold.

There is a particular orientifold projection of Type 0B string theory
for which we show that the crosscap state vanishes when projected onto
physical states. For this model of unoriented strings all contributions
from unoriented worldsheets vanish and the S-matrix 
for the massless tachyon is precisely  the same as in
 oriented Type 0B string theory.  This fits nicely with the matrix
model description of 
this orientifold model, which can be shown to reduce to 
(one-half of) the system of
free fermions describing oriented Type 0B strings. 

The plan of the paper is as follows. In section $2$ we
analyze  unoriented $c=1$ bosonic string theory and exhibit the need to
renormalize the worldsheet theory to cancel divergences arising
from a tachyon tadpole on the $RP^2$ worldsheet. In section $3$
we analyze the matrix model dual and show that it is described by the
dynamics of fermions interacting via a Calogero-Moser
potential. Using the collective field theory description of this
model we show that the tree amplitudes are the same as in the
Hermitian matrix model, as expected from string theory
considerations. We find a double trace deformation of the
Hermitian matrix model which yields the matrix model for unoriented
strings and describe how to write down the non-local worldsheet deformation that
interpolates between oriented and unoriented string theory. In section
$4$ we analyze the thermodynamics of the Calogero-Moser system in an
inverted harmonic oscillator potential using the asymptotic Bethe
ansatz. We show that there is a particle-hole duality in the model,
which implies that the free energies 
of the  dual
$SO$ and $Sp$ orientifold models are related by flipping the sign
of the string coupling constant, $g_s\rightarrow -g_s$, thus
reproducing a well-known property of string theory to all orders in perturbation theory.  Section
$5$ contains a classification of the possible orientifold models of
Type 0B string theory together with the computation of the partition
function. For one of these models we find that the overlap of the
crosscap state with all physical states
vanishes, while the other model has divergences due to a tachyon
tadpole, as in the bosonic string.
In section $6$ we construct the matrix model duals for Type 0B
unoriented strings and show that the
one corresponding to the model with a vanishing crosscap
actually reduces to a system of free fermions, thus confirming the
string theory expectation that only oriented worldsheets
contribute to all orders in perturbation theory.
 The other matrix model is the same as for the
unoriented bosonic string theory, except that now both
sides of the inverted harmonic oscillator
potential are filled. Section $7$ describes possible orientifold projections
of Type 0A string theory and  the computation of their partition
function. Section $8$ analyzes the matrix model duals of Type 0A orientifolds; we show
that all these models reduce to integrable systems of the Calogero-Moser type.

\newsec{Unoriented $c=1$ Strings}

The worldsheet description of $D=2$ bosonic string theory is given by a free
boson tensored with the Liouville\foot{As usual, the $b,c$
reparametrization ghosts must also be included.} CFT with $c_L=25$. The
 local dynamics of the Liouville CFT on a worldsheet $\Sigma$ is 
given by
\eqn\liov{
S=\int_{\Sigma} d^2z\left({1\over 2\pi}\partial
\phi\bar{\partial}\phi+\mu
e^{2b\phi}\right),}
where $Q=b+{1\over b}$ and $c_L=1+6Q^2$. Globally, there is a
background ``charge'' 
$-Q\chi$ for $\phi$, where $\chi=2-2g-b-c$ is the Euler characteristic
of the  worldsheet $\Sigma$, $b$ is the number of boundaries, and $c$ is the number of crosscaps. 
The  vertex operators
corresponding to normalizable states $|p\ra$ are given by
\eqn\norm{
V_\alpha=e^{\alpha\phi}\qquad \hbox{with}\ \alpha={Q}+ip,\
p\geq 0.}
When restricting to the zero mode $\phi_0$, the wavefunction for $|p\ra$ --
which satisfies the minisuperspace Schr\"odinger equation -- is given by
\eqn\wavefunct{
\psi_p(\phi_0)={2(\pi\mu/b^2)^{-ip/2b}\over
\Gamma(-ip/b)}K_{ip/b}(2\sqrt{\pi\mu/b^2}e^{b\phi_0});}
in the weak coupling region $\phi_0\rightarrow - \infty$ it
asymptotes to 
\eqn\limi{
\psi_p(\phi_0)=e^{ip\phi_0}+R^{cl}(p)e^{-ip\phi_0},}
where $R^{cl}(p)$ is the semiclassical reflection
amplitude\foot{See \DornAT\DornXN\ZamolodchikovAA\TeschnerRV\ for
details on the exact computation of the reflection amplitude.}. This
asymptotics reveals complete reflection off the 
Liouville potential and gives a rationale for restricting to states with
$p\geq 0$.

In $D=2$ string theory asymptotic states are
created by the vertex operator 
\eqn\onshell{
V_p=e^{ipX^0}V_\alpha,}
and correspond to excitations of a two-dimensional massless scalar field T, known as the
massless tachyon.

$D=2$ bosonic string theory  is invariant under the action of worldsheet
parity $\Omega$, so that one can mode out the theory by the
orientifold group 
 $G=\{1,\Omega\}$. The tachyon field $T$ is invariant under the action
of $\Omega$ 
  but unoriented worldsheets must now  be included
(worldsheets $\Sigma$ where the number of crosscaps $c$ does not
vanish). At leading order in $g_{st}$ only the spherical worldsheet contributes, and 
all scattering amplitudes are
the same as in oriented string theory\foot{This well known string
theory fact will be deduced in the next section from the dual matrix model viewpoint.}.

The leading unoriented Liouville CFT 
contribution comes from an $RP^2$ worldsheet. We first consider the
simplest amplitude, the one point function. 
One can easily compute
the one point function of $V_\alpha$ in the semiclassical regime by
constructing in the minisuperspace approximation the wavefunction for
the crosscap state
$\la C|$ and evaluating the overlap $\la C|p\ra$. The crosscap
condition \CallanPX\PolchinskiTU , when
restricted to the zero mode, forces $\la C|$ to carry zero momentum so
that the semiclassical wavefunction is $\psi_C(\phi_0)=1$. Therefore,
the semiclassical one point function is given by:
\eqn\onepoint{
\la V_\alpha\ra^{cl}_{RP^2}=\la C|p\ra=\int_{-\infty}^\infty d\phi_0\,\psi_C(\phi_0)
\psi_p(\phi_0)={2\over b}(\pi\mu/b^2)^{-ip/2b}\Gamma(ip/b)\cosh\left({\pi
p\over 2b}\right).}
The exact computation can be performed by bootstrap methods and
gives \HikidaBT
\eqn\onepointexact{
\la V_\alpha\ra_{RP^2}=
(\pi\mu\gamma(b^2))^{-ip/2b}{\Gamma(1+ibp)\Gamma(1+ip/b)\over  i
p}\hskip-2pt\left[\cosh\hskip-2pt\left(\hskip-2pt{\pi p\over 2}\hskip-2pt\left(\hskip-1ptb+{1\over
b}\hskip-1pt\right)\hskip-2pt\right)\hskip-2pt+\hskip-2pt\cosh\hskip-2pt\left(\hskip-2pt{\pi 
p\over 
2}\hskip-2pt\left(\hskip-1pt b-{1\over 
b}\hskip-1pt\right)\hskip-2pt\right)\hskip-2pt\right],} 
where $\gamma(x)={\Gamma(x)\over \Gamma(1-x)}$. In the semiclassical
regime $(b\ll 1)$ the answer reduces to the minisuperspace computation
\onepoint. 

When considering $D=2$ string theory, one defines the Liouville
contribution to amplitudes by
 taking the
following limit
\eqn\double{
b\rightarrow 1,\ \mu\rightarrow \infty\qquad\ \hbox{with}\qquad\
\pi\mu\gamma(b^2)=\mu_R,}
which yields finite results. When $c_L=25$, the amplitude \onepointexact\  has
poles at imaginary integer values of the momentum which signals
the fact that we are dealing with an orientifold plane stretched along
the $\phi$ direction, an $O1$-plane. 

In string theory one is interested in the correlators of the on-shell vertex operators
$V_p$. Energy-momentum conservation forces the one-point function to
be evaluated at zero momentum, so that one is computing  the zero
momentum tachyon 
tadpole on $RP^2$. At first sight, the amplitude \onepointexact\ is
divergent in the zero momentum limit due to a pole at $p=0$. Here we 
interpret this singularity in the CFT computation as a conventional
infrared singularity arising from the noncompactness of the target space. The Liouville potential in \liov\ serves as an
infrared cutoff in the strong-coupling region and effectively puts the system in a box of length
$V_\phi=-{1\over 2}\log\mu$. The zero momentum limit 
should be interpreted as inserting the lowest momentum mode in the
box, which is 
given by $p_{min}={\pi\over V_\phi}$. Therefore, the tachyon tadpole is
given by\foot{$C_{RP^2}$ is a numerical coefficient which depends on the precise normalization of the crosscap state.} 
\eqn\tach{ 
\lim_{p\rightarrow 0}\la V_p\ra_{RP^2}={V_{x^0}\over
2\pi}{V_\phi\over 2\pi}\cdot C_{RP^2},} 
which agrees with the result  obtained by performing a
conventional free field theory computation. The fact that the $RP^2$
tadpole 
is captured by free field theory can also be shown by performing the path
integral over the Liouville zero mode \DiFrancescoUD\GuptaFU\
and noticing that when the vertex
operator is at zero momentum, the Liouville interaction can be safely
ignored.

\ifig\Seminearest{Origin of the divergence of scattering amplitudes on
$RP^2$.}{\epsfxsize4in\epsfbox{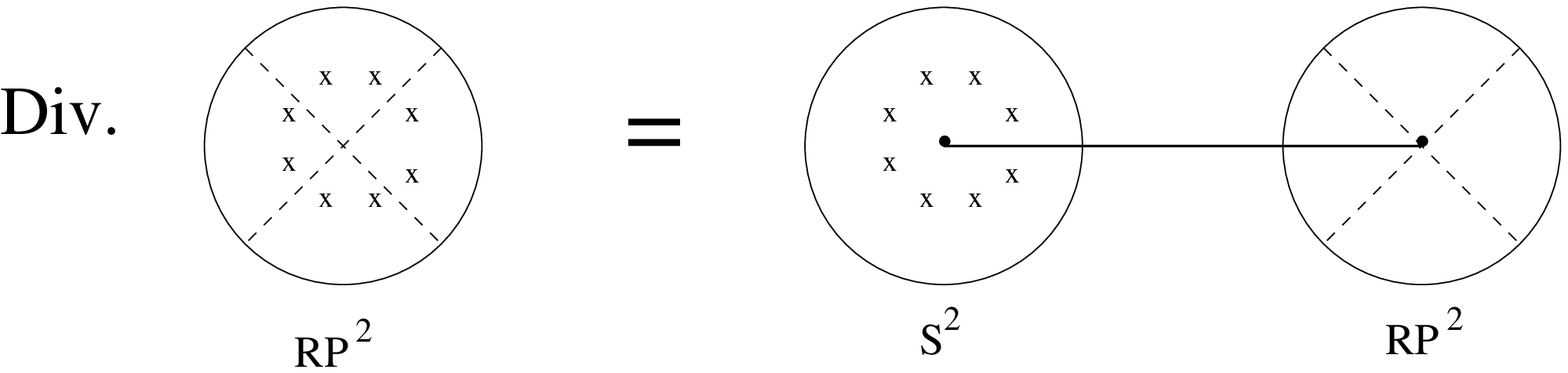}}

A tachyon tadpole on the $RP^2$ worldsheet gives rise to divergences in string
amplitudes, and some care must be taken in interpreting them. In
particular, contributions to the  S-matrix ($n\geq2$-point
functions) arising from the 
$RP^2$ worldsheet are divergent. All these divergences are proportional to the $RP^2$ tadpole.
The region in
the moduli space which yields the divergence is the region where all
vertex operators collide. In this region the worldsheet is conformally
equivalent to a sphere with all vertex operator inserted connected
via a long neck to the $RP^2$ worldsheet. The divergence is therefore
due to the propagation of the zero momentum tachyon and is
proportional to the $RP^2$ tachyon tadpole.

Another example of a divergent string amplitude appears in the
computation of the free energy of the model, which has a torus
and a Klein bottle 
contribution corresponding to the following partition function:
\eqn\part{
Z=\hbox{Tr}\left[\left({1+\Omega\over
2}\right)q^{L_0}\bar{q}^{\bar{L}_0}\right]={1\over 2}Z_{T^2}+{1\over 2}Z_{KB}.}
The torus contribution is well known \BershadskyXB\ and yields
\eqn\torus{
 Z_{T^2}={V_\phi}{1\over 12}\left({R\over
 \sqrt{\alpha^\prime}}+{\sqrt{\alpha^\prime}\over R}\right),}
where $V_\phi$ is the volume of the Liouville direction, which due to
 our conventions is $V_\phi=-{1\over 2}\log\mu$. The first term
 corresponds
to the temperature independent piece of the free energy, while the
 second term is the temperature dependent piece, which can be
 identified with the free energy of a massless scalar field in two
 dimensions\foot{The free energy is given by ${F\over T}=-Z$, where
 $T={\sqrt{\alpha^\prime}\over 2\pi R}$. The temperature dependent
 piece of the free energy of a massless
 scalar field in two dimensions, which is regularization independent,
 is given by $F(T)-F(0)=-{\pi\over 6}V_\phi T^2$.}.

We now compute the Klein bottle amplitude\foot{See also \NakayamaEP\
for the computation of the
zero temperature Klein bottle in the $D=2$ bosonic string .}
by performing the path integral\foot{One can also compute the Klein
bottle by evaluating the crosscap overlap $\int dl \la
C|e^{-lH_c}|C\ra$ using \onepointexact. The naive computation is
divergent, but we interpret the divergence as a harmless infrared (volume)
divergence just as in \tach(the Klein bottle is also proportional
to $1/p_{min}$), and this approach yields precisely the same result
as the path integral.}.
As in
 the torus computation, the contribution of the massive oscillators
cancel the ghost determinant and one is left with a sum over the zero
 modes of the scalar. We must take into account the action of  $\Omega$
 on the momentum and winding zero modes
\eqn\acton{\eqalign{
\Omega |n\ra=|n\ra\cr
\Omega |w\ra=|-w\ra,}}
so that only momentum modes contribute to the trace. Therefore, one
obtains:
\eqn\trace{
Z_{KB}={V_{\phi}\over 2\pi}\int_0^\infty{dt\over
2t^{3/2}}\sum_{n=-\infty}^\infty\exp\left(-\pi t{\alpha^\prime\over
R^2}n^2\right)={V_{\phi}\over 2\pi}{R\over\sqrt{\alpha^\prime}} \int_0^\infty{dt\over 
2t^{2}}\sum_{n=-\infty}^\infty\exp\left(-\pi
{R^2\over\alpha^\prime}{n^2\over t}\right).}
To interpret the amplitude we factorize the Klein bottle into the 
tree channel
$(t=1/l$): 
\eqn\integral{
Z_{KB}={V_{\phi}\over 4\pi}{R\over\sqrt{\alpha^\prime}}\int_0^\infty
dl \left(1+2\sum_{n=1}^\infty \exp\left(-\pi
{R^2\over\alpha^\prime}{n^2 l}\right)\right)={V_{\phi}\over
4\pi}{R\over\sqrt{\alpha^\prime}}\left(\int_0^\infty 
dl+{\pi\alpha^\prime\over 3R^2}\right).}
There is again a divergence arising from propagation of the zero
momentum massless 
tachyon proportional to the square of the $RP^2$ tachyon tadpole.
 The complete partition function of the model is given by:
\eqn\aprtfinal{
Z={V_{\phi}}{1\over
24}\left({R\over \sqrt{\alpha^\prime}}+2{\sqrt{\alpha^\prime}\over
R}\right)+{V_{\phi}\over 4\pi}{R\over\sqrt{\alpha^\prime}}\left(\int_0^\infty
dl\right).}
Therefore, the temperature independent piece of the free energy is
divergent while the temperature dependent piece is
precisely that of a massless two dimensional scalar field, as expected from
space-time considerations. 

All the divergences above can be cancelled via the Fischler-Susskind
mechanism \FischlerCI\FischlerTB.
 This is achieved by adding to the worldsheet Lagrangian a counterterm
corresponding to the zero momentum vertex operator -- which is
precisely the Liouville interaction in \liov\ -- 
 with coefficient proportional to the tadpole
amplitude in $RP^2$. More concretely, we have to renormalize the theory by adding
the worldsheet interaction
\eqn\renorm{
\delta S_{ws}=\log\Lambda\cdot C_{RP^2}\cdot g_s\int d^2z\; e^{2\phi},}
where $\Lambda$ is a UV cutoff on the worldsheet.
The divergence in the $n$-point function on $RP^2$ is now cancelled by
the $n$-point function on the
sphere with one insertion of \renorm, denoted by $\otimes$,
while the Klein bottle
divergence is cancelled by the one-point function of \renorm\ on the
$RP^2$ worldsheet. In this way one can construct finite string
amplitudes in perturbation theory.
\ifig\Seminearest{Cancelling divergence with a
counterterm.}{\epsfxsize3in\epsfbox{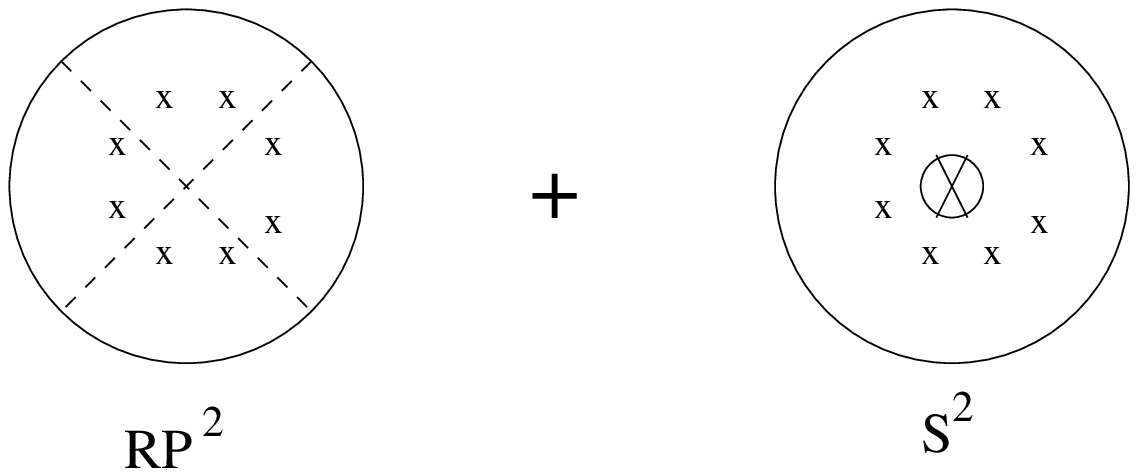}} 
\noindent

\newsec{Unoriented $c=1$ Matrix Model}

\subsec{Derivation of the matrix model}

The dual matrix model is conjectured to be given by the
worldvolume action of $N$ unstable D0-branes in the unoriented
model. The basic idea is that once the tachyon mode on the D0-branes
condenses, one is left with the original closed string background, and
therefore the D0-branes should capture the physics around the tachyon
vacuum.

{}From the work of ZZ \ZamolodchikovAH, it is known
that the degrees of freedom of a D0-brane are described by a Hermitian tachyon field
$M$ and a non-dynamical gauge field $A$.
 In order to determine the worldvolume
theory on a collection of $N$ D0-branes in the orientifold model,
one must demand invariance of the open string vertex operators
under the action of $\Omega$. The result of the projection depends  
on the type
of orientifold projection under study (depending on the sign of the
crosscap state $|C\ra$), and is given by
\eqn\matrices{\eqalign{
O1^-:&\ M^t=M\qquad A^t=-A\cr
O1^+:&\ JM^tJ=-M\qquad JA^tJ=A,}}
where $J$ is the $Sp(N)$ invariant tensor. In the first model $M$ is real and
symmetric and $A$ belongs to the $SO(N)$ Lie algebra, while in the
second model $J\cdot M$ is antisymmetric and $A$ belongs to the $Sp(N)$ Lie
algebra.

Thus the effective action of a collection of $N$ D0-branes in
this orientifold model  is given by the following quantum mechanics model
\eqn\quantum{
S=\int dt\ \Tr\left({1\over 2}(D_tM)^2+U(M)\right),}
where $D_tM=\partial_t+i[A,M]$, $U(x)$ is the tachyon
potential, and the Hermitian matrices $M(t)$ and $A(t)$ satisfy the constraints
\matrices.  

\subsec{Reduction to the eigenvalues}

To analyze the dynamics of the system it is convenient to fix a gauge.
In all three cases ($U(N)$, $SO(N)$ and $Sp(N)$) we may gauge-fix by
requiring $M$ to be diagonal \Mehta\Brezinetal . In the 
$Sp(N)$ case each eigenvalue occurs twice.  The Faddev-Popov determinant is simply the Jacobian $J$  
for the change of coordinates from $M_{ij}$ to the eigenvalues
$\lambda_i$ and the ``angular''  
variables $U^i_j$, where the matrix $U^i_j$ is unitary, orthogonal, or
symplectic-unitary. If we denote by 
$\Delta$ the Vandermonde determinant of the eigenvalues $\lambda_i$,
given by $\Delta(\lambda)=\prod_{i<j}(\lambda_i-\lambda_j)$,
the Jacobian is given by:
\eqn\vandermondejac{\eqalign{
U(N): &\ J=\Delta^2(\lambda)\qquad \hbox{oriented $c=1$}\cr
SO(N): &\ J=\Delta(\lambda)\qquad\hbox{unoriented $c=1$ with $O1^-$} \cr
Sp(N): &\ J=\Delta^4(\lambda)\qquad\hbox{unoriented $c=1$ with $O1^+$}.}}

The equation of motion for $A$ imposes the Gauss' law constraint and
restricts physical states to belong to the singlet sector, where the
Hamiltonian 
simplifies considerably. The Hamiltonian  in the singlet sector is given by:
\eqn\hamil{
H=-{1\over 2J}\sum_i{d\over d\lambda_i}J{d\over d\lambda_i}
+\sum_iU(\lambda_i).}
It is convenient to eliminate
terms in the Hamiltonian with one derivative by
writing
$\Psi(\lambda)=\sign(\Delta) |\Delta|^{-\alpha/2} f(\lambda)$, where $\alpha=2,1,4$ respectively for the cases in 
\vandermondejac. Then
$f(\lambda)$ is an eigenfunction of the following Hamiltonian:
\eqn\hamilcanvi{
{\tilde H}=-{1\over 2}\sum_i{d^2\over
d\lambda_i^2}+\sum_iU(\lambda_i)+{\alpha\over 
2}({\alpha\over 2}-1)\sum_{i<j}{1\over (\lambda_i-\lambda_j)^2}.}
Since $\Psi(\lambda)$ is symmetric under the exchange any two eigenvalues, $f(\lambda)$
is antisymmetric.
Thus we end up with a system of nonrelativistic fermions in an external potential
interacting via the $1/x^2$ potential. This is known as the Calogero-Moser model.
For the Hermitian matrix model ($\alpha=2$), the coefficient
multiplying the interaction potential vanishes, and we recover the standard result that
the dynamics of the eigenvalues is described by free fermions. We
will focus on the values $\alpha=1,4$, which correspond to the 
quantum mechanics of real symmetric and 
``quaternionic-Hermitian'' matrices \matrices. 

We are interested in studying the dynamics of \hamilcanvi\ in the double
scaling limit. In this limit, the only relevant feature of the
potential $U(x)$ is the coefficient of the quadratic maximum, which
given its interpretation as the tachyon potential, encodes
the mass of the open string tachyon \McGreevyKB\KlebanovKM, so that 
$U(x)=-{1\over 2}x^2$. It is well-known that for many purposes,
 particles in the Calogero-Moser model can be
thought of as free, but satisfying a generalized exclusion statistics (see \PolychronakosSX\ for a review). This means
that the number of particles per level cannot be larger than $\alpha/2$. We will discuss this in more detail 
in Section $4$; for now it is sufficient to say that generalized exclusion statistics implies the existence of
a quasi-Fermi surface, much like in the theory of free fermions. Oscillations of the quasi-Fermi surface can be
described by a collective field \JevickiMB , which is defined as the eigenvalue density:
\eqn\density{
\phi(x,t)=\sum_{i=1}^N\delta(x-\lambda_i(t)).}
One also introduces the canonically conjugate momentum $\Pi(x,t)$, which
satisfies 
\break
$[\phi(x,t),\Pi(y,t)]=i\delta(x-y)$. The Hamiltonian for $\phi(x,t)$ and its conjugate momentum $\Pi(x,t)$ is given by
\AndricJK
\eqn\hamilcoll{
H=\int dx \left(\phi \Pi_{,x}^2+{\pi^2\alpha^2\over
12}\phi^3+{\alpha\over 2}({\alpha\over 2}-1)\phi_H \phi_{,x}+{({\alpha\over
2}-1)^2\over 4}{\phi_{,x}^2\over \phi}+U(x)\phi \right),}
where $\phi_H$ is the Hilbert transform of $\phi$:
\eqn\transform{
\phi_H(x)=\int dy\ \phi(y){P\over x-y},}
and $P$ is the principal part. Furthermore, the field $\phi$ is constrained to
satisfy:
\eqn\const{
\int_{-\infty}^{\infty} dx\ \phi(x)=N.}

In order to determine the leading large N behavior, it is useful to
define 
\eqn\behav{
\phi={\sqrt N} {\hat\phi}, \qquad \Pi=N{\hat\Pi},\qquad x={\sqrt N} {\hat x},}
so that $\int d{\hat x}\ {\hat \phi}({\hat x})=1$ and the effective Planck constant is $1/N^2$. Simple scaling shows
that the third and fourth terms in \hamilcoll\ are
subleading in the large $N$ expansion, and that 
the remaining terms are precisely those  of the collective field
theory for the Hermitian matrix model (see e.g. \JevickiQN), and are
proportional to $N^2$, as is typical in the tree level
approximation. At leading order in $1/N$ expansion the only difference between the case $\alpha=1$ 
and the general case is the $\alpha$-dependence of the $\phi^3$ term. It can be absorbed into
an additional rescaling of $\phi,\Pi$ and $x$. This shows 
that unoriented string 
amplitudes at tree level are the same as in the oriented model,
as expected from string theory considerations.

\subsec{Connecting oriented and unoriented strings}

One can view the Calogero-Moser system in an inverted harmonic potential
as a perturbation of the usual free-fermion system by an interaction
term:
\eqn\Hintterm{
H_{int}\sim\int dx\ dy\ \frac{\hrho(x)\hrho(y)}{|x-y|^2}.
} 
where $\hrho(x)=\psi^\dagger(x)\psi(x)$ is the fermion-density operator.
Because
of a short-distance singularity in the potential, a regularization is
required
when doing perturbation theory with this Hamiltonian.

Since free fermions in an inverted harmonic potential are equivalent to
the oriented
$c=1$ bosonic string, we can regard the Calogero-Moser system in an
inverted harmonic
potential as a deformation of the latter. In other words, the
unoriented $c=1$ string
is a deformation of the oriented $c=1$ string!

Let us determine the infinitesimal form of this deformation. It is
convenient to switch
to  momentum space in the $x$-direction. In the language of matrix
quantum mechanics,
the fermion-density operator $\tilde\rho(q,t)$ is:
\eqn\trhoee{
{\tilde \rho} (q,t)=\hbox{Tr}\ e^{iqM(t)}=W_{iq}(t).}
This is known as a macroscopic loop operator (with imaginary length
$iq$). Thus turning
on an interaction potential $V(q)$ between the fermions is equivalent to
adding
a double-trace operator to the matrix model Hamiltonian:
\eqn\Hintmatrix{
H_{int}=\int \frac{dq}{4\pi} W_{iq}(t)W_{-iq}(t)V(q).
}
In the Calogero-Moser case $V(q)\sim |q|$.

Double-trace deformations of the Hermitian matrix model lead to
non-local interactions on
the worldsheet \dtrone\dtrtwo. Such deformations have recently been studied in the
context of AdS/CFT
correspondence \AharonyPA. To find the explicit form of the non-local interaction, one has to expand the macroscopic loop operator as a linear combination of the tachyon vertex operators as in \MooreSeiberg\ . Schematically, this gives the following deformation of the action:
\eqn\nonloc{
\delta S= \int dp V_p V_{-p} K(p),
}
where $K(p)$ is some complicated function whose explicit form is not
very illuminating and $V_p$ is the tachyon vertex operator. 
The conclusion is that one can deform the oriented $c=1$ string into the
unoriented one
by turning on a nonlocal interaction in the worldsheet theory. 

\newsec{Thermodynamics of the Calogero-Moser system in an inverted harmonic potential}

\subsec{Asymptotic Bethe ansatz: a brief review}

We now turn to the
thermodynamic properties of the quantum Calogero-Moser model with the goal
of comparing with the computations in section $2$.
The key feature of the Hamiltonian \hamilcanvi\ is the existence of $N$ independent
integrals of motion whose pairwise 
commutators vanish \CMintegrone\CMintegrtwo . This means that the system is completely
integrable and enables one
to compute its thermodynamic properties exactly. The usual way of
doing this is  
through asymptotic Bethe ansatz. 

Let us recall how this method works in the case when there is no external potential.
Complete integrability implies that the many-body S-matrix is elastic and factorizes into a product
of two-body S-matrices describing pairwise collisions. Imagine putting the
system in a large box $0<x<L$ and let us take the coordinates of all
particles but the $i^{\rm th}$ one 
to be far from the boundaries of the box. Then the wavefunction
$f(x_1,\ldots,x_N)$, regarded 
as a function of $x^i$, is given by plane-waves near the left boundary
of the box: 
\eqn\Psione{
f(x_i)\sim\sin(p_i x_i),\quad x_i\simeq 0.
} 
Then near the right boundary of the box we have:
\eqn\Psitwo{
f(x_i)\sim\exp(i p_i x_i+i\phi_i^+)-\exp(-ip_i x_i-i\phi_i^-).
}
Here $\phi_i^+$ is the total phase shift resulting from 
the $i^{\rm th}$ particle
moving to the right with momentum $p_i$ and colliding successively
with the other $N-1$ particles, 
while $\phi_i^-$ is the total phase shift resulting from the $i^{\rm
th}$ particle  
moving to the left with momentum $-p_i$ and colliding again with the
same particles. 
In other words
\eqn\phipm{
\phi_i^+=\sum_{j\neq i} \phi(p_i-p_j),\quad \phi_i^-=\sum_{j\neq i}
\phi(p_i+p_j), 
}
where $\phi(p)$ is the 2-body phase-shift. Requiring the vanishing of the
wavefunction on the boundary $x=L$, we get the Bethe equations:
\eqn\Betheone{
2p_i L+\sum_{i\neq j} \phi(p_i-p_j)+\sum_{i\neq j}\phi(p_i+p_j)=2\pi
n_i\quad n_i\in\ZZ.
}
This system of equations 
determines the allowed values of $p_i$. Note that $p$ and $-p$ are
physically indistinguishable, so 
we may assume that all $p_i$ are positive. In the thermodynamic limit
$N\rightarrow\infty$ the 
allowed values of momenta are very closely spaced, and we may
introduce the level-density:
\eqn\density{
\rho(p_i)=\frac{\Delta n_i}{\Delta p_i}.
}
For free particles, this density is equal to $L/\pi$. We may formally continue $\rho(p)$
to negative values of $p$ as an even function. The Bethe equations imply a
linear integral equation for $\rho(p)$ which takes account of the interactions:
\eqn\Tzero{
2\pi\rho(p)=2L+\int_{-\infty}^{+\infty} \phi'(p-q) \rho(q) dq.
}
In the derivation it was assumed that all successive energy levels are occupied by particles,
so that the energy 
\eqn\Etot{
E=\sum_i p_i^2=\int_0^\infty p^2 \rho(p) dp
}
is minimized. For non-zero temperature we are supposed to minimize the
free energy instead. Then we must allow holes (unfilled levels) in the
particle distribution. 
Introducing the density of holes $\rho_h(p)$, we get the following equation:
\eqn\Tnotzero{
2\pi(\rho(p)+\rho_h(p))=2L+\int_{-\infty}^{+\infty} \phi'(p-q)\rho(q) dq.
}
This must be supplemented by a relation between $\rho$ and $\rho_h$
obtained by minimizing 
the free energy. The latter is given as a functional of $\rho$ and $\rho_h$:
\eqn\FBethe{
F=E-\mu N -TS=\int_0^\infty \{(p^2 -\mu)\rho(p)-T\left[(\rho+\rho_h)\log(\rho+\rho_h)-\rho\log\rho-\rho_h\log\rho_h\right]\} dp.
} 
In general, this leads to a non-linear integral equation for $\rho$. Once its solution is found, one can find
the free energy by substituting $\rho$ and $\rho_h$ into \FBethe.

If instead of a simple impenetrable box we have a boundary potential at one end, and if this potential
preserves the integrability of the system, then the Bethe equations
are modified 
as follows:
\eqn\Bethetwo{
2 p_i L+\phi_b(p_i)+\sum_{i\neq j} \phi(p_i-p_j)+\sum_{i\neq
j}\phi(p_i+p_j)=2\pi n_i,\quad n_i\in\ZZ. 
}   
Here $\phi_b(p)$ is the phase-shift describing reflection off the
boundary potential. 
In the thermodynamic limit we can replace this equation by a linear integral
equation for $\rho$ and $\rho_h$:
\eqn\lineqbdry{
2\pi(\rho(p)+\rho_h(p))=2L+\phi_b'(p)+\int_{-\infty}^{+\infty} \phi'(p-q)\rho(q) dq.
}

\subsec{Asymptotic Bethe ansatz for the Calogero-Moser system in an inverted harmonic potential}

In the case of interest to us (Eq. \hamilcanvi) the role of the boundary
potential is played by the inverted 
harmonic oscillator potential $U(x)=-{1\over 2}x^2$. This entails
certain modifications in 
the arguments. 
Firstly, the inverted harmonic potential allows for tunneling to the
other side of the 
potential; however, if we are interested in the semiclassical
asymptotics (corresponding 
to the double-scaling limit in the matrix model), then the tunneling is
exponentially small, and 
can be neglected. Secondly, and more importantly, the potential does
not tend to a constant 
at infinity, and the effect of the external potential on the
asymptotic behavior of the wavefunction does not 
amount to an $x$-independent phase shift. Since the external potential
breaks translational 
invariance even at infinity, one has to label one-particle states by energy $\eps$
rather than momentum. Then at large distances the wavefunction
behaves as follows: 
\eqn\Psitwo{
f(x_i)\sim\hskip-2pt \frac{1}{\sqrt x_i}\hskip-2pt\left[\exp(-ix_i^2/2-i\eps_i\log x_i-\phi_-(\eps_i))\hskip-2pt -\hskip-2pt\exp(ix_i^2/2+i\eps_i \log x_i + 
\phi_b(\eps_i)+\phi_+(\eps_i))\right].}
Here $\phi_b(\eps)$ can be deduced by solving the Schr\"odinger equation for a single particle
in an inverted harmonic potential and is given by \MooreSF:
\eqn\bdryrefl{
\phi_b(\eps)=\frac{\pi}{2}+\eps\log 2+\arg
\Gamma\left(\frac{1}{2}-i\eps\right). 
}
Note that in the derivation of the above equation we have implicitly
assumed that the S-matrix 
factorizes into the product of two-body S-matrices and the boundary
S-matrix. This can be 
justified using the complete integrability of the system.

The Bethe equations now look as follows
\eqn\Bethemod{
L^2+2\eps_i\log L+\phi_b(\eps_i)+2\sum_{i\neq j}
\phi(\eps_i-\eps_j)=2\pi n_i,\quad n_i\in \ZZ. 
}
The corresponding integral equation for particle and hole densities is:
\eqn\linearmod{
2\pi (\rho(\eps)+\rho_h(\eps))=2\log L+\phi'_b(\eps)+2\int_{-\infty}^{+\infty} \phi'(\eps-\eps')\rho(\eps') d\eps'.
}

For the Calogero-Moser interaction potential $U={l(l-1)\over
(\lambda_1-\lambda_2)^2}$, the two-body phase 
shift is given by\foot{In this section we set $l={\alpha\over 2}$.} (see e.g. \PolychronakosSX):
\eqn\CMshift{
\phi(\eps)=\pi (1-l) \sign(\eps).
}
Therefore the integral equation for $\rho$ and $\rho_h$ becomes a simple algebraic relation:
\eqn\CMlin{
2\pi(\rho(\eps)+\rho_h(\eps))=2\pi (1-l)\rho(\eps)+2\log L+\phi'_b(\eps).
}
In other words:
\eqn\CMlintwo{
\rho_h(\eps)+l\cdot\rho(\eps)=\frac{\log L}{\pi} +\frac{1}{2\pi}\phi'_b(\eps).
}
The right-hand side of this equation is precisely the one-particle density
of states for non-interacting fermions in an inverted harmonic
potential (with a wall at $x=L$). 
Note that it is natural to change variables from $x$ to $\tau=\log x$;
then $\log L$ can 
be interpreted as the size of the box in the $\tau$ coordinate.
For $l=1$ the equation \CMlintwo\ simply says that the sum of the
particle and hole 
densities must add up to the density of available one-particle
states. The remarkable 
feature of the Calogero-Moser potential is that its entire effect is to replace $\rho$
with $l\cdot \rho$. 

Put in a more intuitive way, this means that we may treat Calogero-Moser particles as non-interacting and
obeying a generalized exclusion principle: the distance between occupied energy levels must be
at least $l$ \genstatone\genstattwo\PolychronakosSX . The cases $l=0$ and $l=1$ correspond to bosons and
fermions, respectively. 
For non-integer $l$ the generalized exclusion principle makes sense only once one averages over 
many energy levels. As shown in the previous section, the symmetric
matrix quantum mechanics model corresponds to $l=1/2,$ which is
half-way between bosons 
and fermions 
(``semions''), and is dual to the $O1^-$ orientifold. 
The ``quaternionic-hermitian''  matrix quantum mechanics model
 corresponds to $l=2$, which means that there should be at least one hole
between any two occupied levels (``doubly-fermions''), and is dual to
the $O1^+$ orientifold . To summarize, we can determine all thermodynamic quantities of
the Calogero-Moser system in the inverted harmonic potential by treating the particles as non-interacting but obeying generalized
exclusion statistics.

Now we are ready to compute the free energy. We express $\rho_h$ in
terms of $\rho$ and then minimize the functional \FBethe. The result is:
\eqn\FCM{
F=-T\int_{-\infty}^\infty \rho_0(\eps) \log \left(1+ {n(\eps)\over {1-l\cdot n(\eps)}}\right) d\eps.
}
Here $\rho_0(\eps)$ is the one-particle level-density in the presence
of the inverted harmonic potential \MooreSF:
\eqn\rhonot{
\rho_0(\eps)={1\over\pi}\log (L\sqrt 2)-{1\over 2\pi}\Re\ \psi\left(\frac{1}{2}-i\eps\right),\quad 
\psi(u)=\frac{d}{du}\log\Gamma (u).
}  
The effective filling factor $n(\eps)$ is determined from the following algebraic equation \genstattwo:
\eqn\neff{
\frac{n(1-(l-1)n)^{l-1}}{(1-l\cdot n)^l}=\exp\left(\frac{\mu-\eps}{T}\right).
}
For $l=1$ the latter equation implies that $n(\eps)$ is the usual Fermi distribution and $F$ is the free energy of 
free fermions in an inverted
harmonic potential:
\eqn\FermiF{
l=1: \quad n={1\over {e^{{\eps-\mu}\over T}+1}},\quad F=-T\int_{-\infty}^\infty
\rho_0(\eps)\log\left(1+e^{{\mu-\eps}\over T}\right) d\eps.
} 
This integral representation for $F$ is divergent, but if we differentiate it with respect to $T$, we
get a convergent integral, which can be expanded in powers of
$1/\mu$. This procedure reproduces the temperature-dependent 
part of the free energy of the Hermitian matrix model \GrossUB:
$$
F(T,\mu)-F(0,\mu)=\frac{\pi T^2}{12}\log\mu+O(\mu^{-2}).
$$

For $l=1/2$ (``semions'') we get:
\eqn\semiF{
l={1\over 2}: \quad n(\eps)=\frac{1}{\sqrt{\frac{1}{4}+\exp\left(\frac{2(\eps-\mu)}{T}\right)}},
\quad F=-T\int_{-\infty}^\infty \rho_0(\eps)\log\left(\frac{1+\frac{n}{2}}{1-\frac{n}{2}}\right) d\eps.
}
For $l=2$ (``doubly-fermions'') we get:
\eqn\doubleFermiF{
l=2: \quad n(\eps)=\frac{1}{2}\left(1-\frac{1}{\sqrt{1+4\exp\left(\frac{\mu-\eps}{T}\right)}}\right),\quad
F=-T\int_{-\infty}^\infty \rho_0(\eps)\log\left(\frac{1-n(\eps)}{1-2n(\eps)}\right) d\eps.
}
The integrals are again divergent, but their temperature-dependent part is convergent. Expanding the
integrands in powers of $1/\mu$, we find:
\eqn\semiFexp{
l={1\over 2}: \quad F(T,\mu)-F(0,\mu)=\frac{\pi T^2}{12}\log\mu+\frac{\zeta(3) T^3}{4\pi\mu}+O(\mu^{-2}),
}
and 
\eqn\doubleFermiFexp{
l=2: \quad F(T,\mu)-F(0,\mu)=\frac{\pi T^2}{12}\log\mu-\frac{\zeta(3) T^3}{2\pi\mu}+O(\mu^{-2}).
}
Note that the leading-order terms are of order $\log\mu$ and are the
same in all three cases 
($l=1/2,\ 1,\ 2$). This is to be expected, as the physics can be
described in all three cases by 
a collective field, and to leading order in $1/N$ the collective field
Lagrangians are identical, as shown in the last section. In fact, 
the $\log \mu$ piece in the free energy is simply the free energy of a
massless two dimensional  
scalar field 
in a box of length $V_\phi=-\frac{1}{2}\log\mu$. This is in agreement
with the string theory result \aprtfinal, which says that the only bulk degree 
of freedom is a massless tachyon. Note also that while for $l=1$ only even powers of $\mu$ appear in the
expansion of $F(T,\mu)$, for $l=1/2$ and $l=2$ all positive powers of $\mu^{-1}$ contribute. This
agrees with the identification of $\mu$ with $1/g_{st}$. In paticular,
the subleading piece in the temperature-dependent part of $F(T,\mu)$ comes from an unorientable worldsheet
with Euler characteristic $-1$.

The temperature-independent part (i.e. the ground state
energy) can be determined from:
\eqn\Egreq{
\frac{\partial E}{\partial \mu}=\mu\rho(\mu)={1\over l}\mu\rho_0(\mu).
}
Thus the ground-state energy is simply $1/l$ times the well-known
free-fermion result: 
\eqn\Eground{
E=F(0,\mu)=\frac{1}{2\pi l}\left(-\frac{1}{2}\mu^2\log\mu+\frac{1}{24}\log\mu-\sum_{m=1}^\infty \frac{(2^{2m+1}-1)|B_{2m+2}|}{8m(m+1)
(2\mu)^{2m}}\right).
}

\subsec{Particle-hole duality for the Calogero-Moser system}

In the Hermitian matrix
quantum mechanics the expansion of the free energy in
powers of $1/\mu$ contains only 
even powers. {}From the string theory point of view this follows from
the identification 
of $1/\mu$ with the string coupling and the fact that only oriented
worldsheets contribute. 
{}From the free-fermion point of view, this is a consequence of the
particle-hole duality, which exchanges 
$\rho$ and $\rho_h$ and takes $\mu$ to $-\mu$. Technically,
particle-hole duality holds to all orders in $1/\mu$ because 
the one-particle density of states $\rho_0(\eps)$ is even. 

It turns out that for $l\neq 1$ there is a generalized particle-hole duality 
which maps the Calogero-Moser system with $l=l_0$ into a
Calogero-Moser system with $l=1/l_0$.  
More precisely, we claim that the following relation holds for the
free energy of the Calogero-Moser 
particles in an inverted harmonic potential:
\eqn\phduality{
F\left(T,\mu ;\ l=\frac{1}{l_0}\right)=l_0^2\ F\left( \frac{T}{l_0},-\mu ;\ l=l_0\right)+\ {\rm terms\ analytic\ in}\ \mu
}
To derive this relation, we define:
\eqn\phdone{
\trho=l\cdot\rho_h,\quad \trho_h=l\cdot \rho,\quad \tl={1\over l},\quad \tT=l\cdot T, \quad \tmu=-\mu .
}
If $\rho$ and $\rho_h$ satisfy $\rho_h+l\cdot\rho=\rho_0$, then $\trho$ and $\trho_h$ satisfy $\trho_h+\tl\cdot\trho=\rho_0$.
Furthermore, one can easily see that the free energy functional evaluated on the tilded quantities is $l^2$ times
the free energy functional of the untilded quantities, up to a (divergent) term which is independent of $\rho$ and
and $\rho_h$ and is linear in $\mu$. This implies the desired result.

For $T=0$ generalized particle-hole duality says that:
\eqn\Egrthree{
E\left(\mu;\ l={1\over l_0}\right)=l_0^2\ E(-\mu;\ l=\ l_0) \ +\ {\rm terms\
analytic\ in}\ \mu. 
}
We already know that $E(\mu;l)$ is $1/l$ times a function of $\mu$
which is even up to analytic terms.  
Hence the relation Eq.~\Egrthree\ is satisfied. Particle-hole duality also explains
why the logarithmic terms in Eq.\semiFexp\ and Eq.\doubleFermiFexp\ are identical, while the
terms of order $\mu^{-1}$ differ by a factor $-2$.

For $l_0=2$ the generalized particle-hole duality has a simple interpretation in terms of dual string theories: it essentially
says that the partition functions of the string theories corresponding to
$l=1/2$ and $l=2$ are related by the substitution $g_{st}\rightarrow -g_{st}$, i.e. by
changing the sign of the contribution from worldsheets with odd Euler characteristic. 
To see this more clearly, we first rewrite the relation Eq.~\phduality\ in terms of the partition functions of the
Calogero-Moser systems:
\eqn\phdZ{
\log Z_{CM}(T,\mu;\ l=1/2)=2\log Z_{CM}\left({T\over 2},-\mu;\ l=2\right)\ +\ {\rm terms\ analytic\ in}\ \mu .
}
Next we note that because in the quaternionic-Hermitian case each eigenvalue of $M$ appears twice, the Hamiltonian of the
quaternionic-Hermitian matrix quantum mechanics is twice the Hamiltonian of the corresponding Calogero-Moser
system. This implies that the definitions of temperature in the Calogero-Moser system and the matrix quantum mechanics
differ by a factor of two. Hence the partition functions of the matrix quantum mechanics are related as follows:
\eqn\phdZtwo{
\log Z^{SO}_{MQM}(T,\mu)=2\log Z^{Sp}_{MQM}(T,-\mu)\ +\ {\rm terms\ analytic\ in}\ \mu .
}
On the other hand, the partition functions in string theory satisfy
\eqn\ZstSoSp{
Z_{st}^{SO}(T,g_{st})=Z_{st}^{Sp}(T,-g_{st}).
}
This equation agrees with Eq.~\phdZtwo\ if we identify $Z_{st}^{SO}$ with $\log Z^{SO}_{MQM}$
and $Z_{st}^{Sp}$ with $2\cdot \log Z^{Sp}_{MQM}$. The normalization factor $2$ in the last equation is needed
to make the leading term in the temperature-dependent part of $F^{Sp}_{MQM}(T,\mu)$ be equal to the free energy of
a massless two-dimensional  scalar field 
in a box of length $-{1\over 2}\log\mu$.

\newsec{Unoriented ${\hat c}=1$ Type 0B Strings}

The worldsheet description of $D=2$ superstring theory is
given by a  free ${\cal N}=1$ superfield tensored with ${\cal N}=1$
super-Liouville with ${\hat c_L}=9$.  ${\cal N}=1$
super-Liouville is defined by an ${\cal N}=1$ superfield $\Phi$ with
components $(\phi,\psi)$ and 
a superpotential $W=\mu e^{b{\Phi}}$. The GSO projection giving rise to
Type 0B string theory is given by:
\eqn\GSO{\eqalign{
\hbox{NS-NS}:&\ {1\over 2}(1+(-1)^{F+\tilde F})\cr
\hbox{RR}:&\ {1\over 2}(1+(-1)^{F+\tilde F}).}}
The spectrum of the model consists of a massless tachyon $T$ arising form the
NS-NS sector and an additional massless scalar $C$ arising from the RR sector. 

Due to the non-chiral GSO
projection, the theory is
manifestly invariant under the action of worldsheet parity
${\Omega}$. We define the action of ${\Omega}$ on the fermions and
on the spin fields as follows:
\eqn\actionon{\eqalign{
&\Omega\cdot\psi={\bar \psi},\
\Omega\cdot{\bar\psi}=-{\psi}\Longrightarrow\ 
\Omega\cdot({\bar \psi}\psi)={\bar \psi}\psi,\cr
&\Omega\cdot S^\alpha={\bar S^\alpha},\
\Omega\cdot{\bar S^\alpha}=S^\alpha\Longrightarrow\ \Omega\cdot({\bar
S^\alpha}\otimes S^\beta)=-{\bar
S^\beta}\otimes S^\alpha.}}

The action of $\Omega$ can be combined with the action of
other $\ZZ_2$ symmetries to yield other consistent orientifold
projections. The theory is  invariant under $(-1)^{F^L_s}$,
where $F^L_s$ is left-moving space-time fermion number, so that one
can consider  
modding out the theory by the following two orientifold groups\foot{In
ten dimensions one can also consider the model (see e.g. \BergmanKM\DudasWD)
obtained by modding out
by the orientifold group
$G=\{1,\Omega(-1)^F\}$. In $D=2$, $(-1)^F$ is not a symmetry since it
flips the sign of the term ${\bar \psi}\psi e^{b\phi}$ in the
super-Liouville action.}:

\eqn\modelos{\eqalign{
\hskip-190pt 1)\;G_1&=\{1,\Omega\}\cr
\hskip-190pt 2)\;G_2&=\{1,\Omega(-1)^{F^L_s}\}.}}

We now analyze the physics of these two models.

\vskip .3in
\noindent
$\bullet\ ${\it Modding Type 0B by $G_1=\{1,\Omega\}$}
\vskip .2in
Given the action of $\Omega$ in \actionon\ it follows that the only state
surviving the projection is $T$. Therefore the space-time physics of
this orientifold model is described by a massless scalar field $T$.

To understand better the spectrum of the model we now compute the
partition function, which is given by:
\eqn\partob{
Z=\hbox{Tr}_{NS-NS\oplus RR}\left[\left({1+\Omega\over
2}\right)\left({1+(-1)^{F+\tilde F}\over
2}\right)q^{L_0}\bar{q}^{\bar{L}_0}\right]={1\over 2}Z_{T^2}+{1\over
2}Z_{KB}.}  
The torus contribution has recently been computed \DouglasUP\
and is given by
\eqn\torusb{
 Z_{T^2}={V_\phi}{1\over 12}\left({R\over
 \sqrt{\alpha^\prime}}+2{\sqrt{\alpha^\prime}\over R}\right),}
where $V_\phi$ is the volume of the Liouville direction.

In computing the Klein bottle contribution we note that the
contribution from $\Omega$-even states, which are the only states  contributing
to the trace,  are automatically
$(-1)^{F+\tilde F}$-even. Therefore we only need to compute
\eqn\kelin{
Z_{KB}=\hbox{Tr}_{NS-NS\oplus
RR}\left[\Omega q^{L_0}\bar{q}^{\bar{L}_0}\right].} 
Now we note that the oscillator contribution from the free superfield
and super-Liouville are  precisely cancelled by the
contribution from the ghosts and superghosts determinants in each
sector. Therefore, the Klein bottle is determined by the action of 
the orientifold group on the zero modes. Combining \acton\ with the fact that
$\Omega\cdot|0\ra_{NS-NS}= |0\ra_{NS-NS}$, it follows that the Klein
bottle contribution {\it vanishes}. Therefore, the partition function
of the model is given by
\eqn\partimodelo{
Z={1\over 2}Z_{T^2}={V_\phi}{1\over 24}\left({R\over
 \sqrt{\alpha^\prime}}+2{\sqrt{\alpha^\prime}\over R}\right),}
whose temperature dependent contribution is precisely that of a
 massless two-dimensional field, as expected from space-time
 considerations. 

One can extract further information about the contribution from
unoriented surfaces by analyzing the crosscap state $|C\ra$, which
describes the closed string state produced by adding a crosscap to a
worldsheet. $|C\ra$ must be invariant under the gauge symmetries of the
model, in particular it has to be GSO invariant. Moreover, it must
reproduce the Klein bottle loop amplitude \partob\ when factorizing
the computation of its norm\foot{When computing the norm one must insert the damping factor $e^{-l H_c}$.} into 
the loop channel:  
\eqn\facto{
\la C|e^{-l H_c}|C\ra=\hbox{Tr}_{NS-NS\oplus
RR}\left[\Omega\left({1+(-1)^{F+\tilde F}\over
2}\right)q^{L_0}\bar{q}^{\bar{L}_0}\right].}
Since the right-hand side vanishes, so does the left-hand side. Inserting a complete
set of physical states, we see that the overlap (regularized by $e^{-l H_c}$) of the crosscap state with any
physical state must vanish. 

One can see the vanishing of the overlap of the crosscap with 
the physical states more explicitly as follows.
By analyzing the effect of the twists when factorizing from the loop
channel to the tree channel, one finds that $|C\ra$ only has contributions
from the NS-NS 
sector
\eqn\crosscap{
|C\ra={1\over \sqrt{2}}(|NSNS,+\ra-|NSNS,-\ra),}
where the signs $\pm$ label the spin structure on $RP^2,$ and $|NSNS\ra$
solves the crosscap conditions \CallanPX\PolchinskiTU. The absence of
the RR contribution is to be expected, as there are no physical states
in the RR sector.
In the NS-NS sector the only physical state
is the tachyon, which is the ground state.
Since $|NSNS,+\ra$ and $|NSNS,-\ra$ differ only by oscillator terms, it follows that the projection of 
$|C\ra$ on the tachyon state vanishes. 

The vanishing of the overlap between the crosscap and the physical states implies that the contribution of 
all unorientable worldsheets to the S-matrix also vanishes.
At first sight, this seems rather strange. However, we will see that this is precisely what the matrix model
predicts. Moreover, it predicts that the tachyon S-matrix in this model is equal to the tachyon S-matrix of the oriented
$\hat{c}=1$ Type 0B string theory. 

\vskip .3in
\noindent
$\bullet\ ${\it Modding Type 0B by $G_2=\{1,\Omega(-1)^{F^L_s}\}$}
\vskip .2in
The effect of the extra $(-1)^{F^L_s}$ projection results in important
differences in the spectrum. In the NS-NS sector $(-1)^{F^L_s}$ acts
trivially, so that $T$ again survives the projection. In the RR
sector $(-1)^{F^L_s}$ acts with an extra minus sign so that $C$ is now
invariant under the orientifold projection, and the spectrum is the same
as in Type 0B oriented string theory. The dynamics, however, is rather
different.

We now proceed to compute\foot{We will be rather brief in this section
since up to 
some important minus signs the computations are the same as in the
previous model.} 
the partition function of the model, which
can be obtained from \partob\ by substituting $\Omega\rightarrow
\Omega(-1)^{F^L_s}$. The toroidal contribution is  the same and
is given by \partimodelo. The Klein bottle contribution is very
different since now both the NS-NS and RR ground states
are even under $\Omega(-1)^{F_s}$, so that their contributions
add instead of cancelling each
other as in the previous model. Therefore, the crosscap state for this
orientifold model is given by:
\eqn\crosscapa{
|C\ra={1\over \sqrt{2}}(|NSNS,+\ra+|NSNS,-\ra).}

It follows that the Klein bottle
contribution for this orientifold is twice that of the bosonic
orientifold \integral . As in the bosonic string orientifold, the
model has a divergence as a result of a massless tachyon tadpole on the
$RP^2$ worldsheet, which is the same as in the bosonic model up to an
extra factor of $\sqrt{2}$. We cancel the Klein bottle divergence
via the
Fischler-Susskind mechanism as in the bosonic model by adding to the
worldsheet a counterterm 
proportional to the super-Liouville interaction operator. After cancelling
this divergence the resulting partition function is:
\eqn\finalc{
Z^{ren}={V_{\phi}}{1\over
24}\left({R\over \sqrt{\alpha^\prime}}+4{\sqrt{\alpha^\prime}\over
R}\right).}
We note that the temperature dependent contribution is that of {\it
two} massless two-dimensional fields, as expected from space-time
considerations. 

\newsec{Unoriented ${\hat c}=1$ 0B Matrix Model}

Following the discussion in section 3, it is natural to conjecture
that the matrix model description of the Type 0B orientifold models
is given by the worldvolume action of $N$ unstable D0-branes. 

The spectrum of open strings on D0-branes in Type 0B string theory is
the same \AhnEV\FukudaBV\  as in the bosonic string, that is, there is
a tachyon mode 
$M$ and a non-dynamical gauge field $A$. As in section 3, in order to
determine the matrix model dual for  the Type 0B orientifolds, we must
understand the action of the orientifold group on the open string
spectrum. The vertex operators  for the tachyon and gauge field 
in  Type 0B string theory contain extra fermions compared to the
vertex operators in the bosonic string, which  result in different
projections when modding out by the orientifold group.

\vskip .3in
\noindent
$\bullet\ ${\it Modding Type 0B by $G_1=\{1,\Omega\}$}
\vskip .2in
When computing the action of $\Omega$ on the open string 
tachyon state, one finds that it is odd\foot{The transformation
properties of the tachyon can be extracted from the non-vanishing of
the disk amplitude with a RR vertex operator  and the open string
tachyon and using the known transformation properties of the RR vertex
operator under various symmetries.}. Therefore, the worldvolume
spectrum for this orientifold model is given by (depending on the sign of the
crosscap state) 
\eqn\matricesb{\eqalign{
O1^-:&\ M^t=-M\qquad A^t=-A\cr
O1^+:&\ JM^tJ=M\qquad JA^tJ=A,}}
where $J$ is the usual $Sp(N)$ invariant tensor. For this
orientifold projection,  both the open string tachyon and the gauge
field belong to the
$SO(N)$ or $Sp(N)$ 
Lie algebra. As before, the equation of motion for $A$ just
forces one to consider states in the singlet sector. 

Therefore, the matrix model dual to this orientifold model is given by
\quantum\ with the matrices belonging to the $SO(N)$ or $Sp(N)$ 
Lie algebra and with the potential for the tachyon now being even,
that is $U(x)=U(-x)$. In order to analyze these matrix models, we
go to the eigenvalue basis. As in section $3$ we must compute the
Jacobian of the change of coordinates. 
  By using the fact that the generators of the $SO(N)$ Lie
algebra are antisymmetric and those of $Sp(N)$ are symmetric, one finds
that the Jacobian factors are given by \MyersYD\ :
\eqn\jacob{\eqalign{
O1^-:&\ J=\prod_{i<j}(\lambda_i-\lambda_j)^2(\lambda_i+\lambda_j)^2\cr
O1^+:&\ J=\prod_i
\lambda_i^2\prod_{i<j}(\lambda_i-\lambda_j)^2(\lambda_i+\lambda_j)^2.}}
Therefore, the Schr\"odinger equation that needs to be solved -- in the
singlet sector -- is given by \hamil . As in section 3, we redefine the wavefunction 
$\Psi(\lambda)=\sign(J)|J|^{-1/2}f(\lambda)$ to eliminate terms with a single
derivative. It is now a straightforward computation to show that
$f(\lambda)$ is an eigenfunction of the following Hamiltonian:
\eqn\hamilfree{{\tilde H}=-{1\over 2}\sum_i{d^2\over
d\lambda_i^2}+\sum_iU(\lambda_i).}
Therefore, the solution of this matrix model reduces to analyzing the
dynamics of $N$ free fermions in an inverted harmonic oscillator
potential. This is identical to the result that one gets when one 
rewrites the Hermitian matrix model in terms of its eigenvalues. The
only difference compared to 
that case is that now wavefunctions are either even or odd depending
on the choice of the orientifold projection:
\eqn\symmconst{\eqalign{
O1^-:&\ f(-\lambda_i)=f(\lambda_i)\cr
O1^+:&\ f(-\lambda_i)=-f(\lambda_i).}}
Therefore scattering amplitudes for the $SO(N)$ and $Sp(N)$ Lie algebra valued
matrix models are exactly the same as for the tachyon in the
${\hat c}=1$ oriented Type 0B string theory. As for the free energy, it is
half the free energy of the ${\hat c}=1$ oriented Type 0B string.

This fits perfectly with the conjectured string theory dual, Type 0B
string theory modded out by $G_1=\{1,\Omega\}$. As discussed
in section $4$, this string theory has only contributions from
oriented Riemann surfaces. Moreover, we showed that the free energy is one
half that for oriented ${\hat c}=1$ Type 0B string theory.

\vskip .3in
\noindent
$\bullet\ ${\it Modding Type 0B by $G_2=\{1,\Omega(-1)^{F^L_s}\}$}
\vskip .2in
As before, the matrix model is obtained by projecting the open string
spectrum by $G_2$. The open string tachyon on the D0-branes is
separately odd under 
$\Omega$ and $(-1)^{F^L_s}$. Therefore, the worldvolume spectrum for this
orientifold model is given by:
\eqn\matricesc{\eqalign{
O1^-:&\ M^t=M\qquad A^t=-A\cr
O1^+:&\ JM^tJ=-M\qquad JA^tJ=A.}}
This is precisely the same spectrum as for the $c=1$ bosonic string
orientifold. 

As shown in section $3$, the dynamics of this matrix model is that of
fermions in an inverted harmonic potential interacting via the Calogero-Moser potential
\hamilcanvi. The only difference in solving this model compared to its
bosonic string analog is that when constructing the quasi-Fermi sea
for the dual of the Type 0B orientifold one fills the states symmetrically 
with respect to the tachyon maximum, due to the $\ZZ_2$ symmetry of the open
string tachyon potential \TakayanagiSM\DouglasUP\ . Therefore the free
energy in this case should be twice that of the bosonic string,
in agreement with the computation in section $4$. Further,
all tachyon scattering amplitudes coincide with those for the bosonic string
to all orders in perturbation theory.

\newsec{Unoriented ${\hat c}=1$ Type 0A Strings}

The worldsheet CFT is the same as for two-dimensional Type 0B
string theory. The GSO projection is different and is given by:
\eqn\GSOa{\eqalign{
\hbox{NS-NS}:&\ {1\over 2}(1+(-1)^{F+\tilde F})\cr
\hbox{RR}:&\ {1\over 2}(1-(-1)^{F+\tilde F}).}}
The spectrum of the model is a massless tachyon $T$ arising form the
NS-NS sector and two non-dynamical gauge fields $C_+$ and $C_-$
arising from the RR sector. More precisely, the gauge fields can be
written as $C_{\pm}={1\over \sqrt{2}}(C\pm\tilde C)$, where $C,{\tilde
C}$ are the vacuum states of the $(R+,R-),(R-,R+)$ sectors respectively.
The model has stable D0-branes
which couple\foot{The
other gauge field $C_+$, has no physical D0-brane which couples to it \DouglasUP .} to $C_-$; they play an important role when deriving the
matrix model dual.

The Type 0A GSO projection is non-chiral, so we can construct new
models by modding out by the allowed orientifold groups $G$. The symmetries are the same as in Type 0B. Therefore we can
quotient the theory by the orientifold groups $G_1$ or
$G_2$ in \modelos. 

\vskip .3in
\noindent
$\bullet\ ${\it Modding Type 0A by $G_1=\{1,\Omega\}$}
\vskip .2in
The tachyon mode is invariant under the action of $\Omega$. In the RR
sector $\Omega$ acts by identifying the two RR sectors:
\eqn\identifymodel{
\Omega\cdot C=-{\tilde C}\qquad \Omega\cdot {\tilde C}=-{C}.}
Therefore, only $T$ and $C_-$ survive the orientifold projection. 

The partition function of the model is given by:
\eqn\partooa{\eqalign{
Z&=\hbox{Tr}_{NS-NS}\hskip-3pt\left[\left(\hskip-2pt{1+\Omega\over
2}\hskip-2pt\right)\hskip-3pt\left(\hskip-2pt{1+(-1)^{F+\tilde F}\over
2}\hskip-2pt\right)\hskip-3ptq^{L_0}\bar{q}^{\bar{L}_0}\hskip-1pt\right]
\hskip-3pt+\hskip-3pt\hbox{Tr}_{RR}\hskip-3pt  
\left[\left({1+\Omega\over   
2}\right)\hskip-3pt\left({1-(-1)^{F+\tilde F}\over
2}\right)\hskip-3ptq^{L_0}\bar{q}^{\bar{L}_0}\right]\cr
&={1\over 2}Z_{T^2}+{1\over
2}Z_{KB}.}}
The torus amplitude has been  computed in \DouglasUP\ and is given by:
\eqn\torusa{
 Z_{T^2}={V_\phi}{1\over 12}\left({2R\over
 \sqrt{\alpha^\prime}}+{\sqrt{\alpha^\prime}\over R}\right).}
A simple consequence of the action of $\Omega$ on the RR sectors
 \identifymodel\ is that the RR contribution to the Klein bottle
vanishes. Since the action is off-diagonal, the RR states do not
 contribute to the trace. Therefore, the Klein bottle is given
 precisely by \integral\ and it has a divergent contribution arising
 from a tachyon tadpole on $RP^2$. This can be easily shown by
 analyzing the crosscap state for this model, which is given by
\eqn\factor{
|C\ra=|NSNS,+\ra}
and has a non-trivial coupling to the zero momentum tachyon.
After cancelling the divergence due to the tachyon tadpole  via the
 Fischler-Susskind mechanism as explained in section $2$, the
 partition function is given by:
\eqn\partioafinal{
Z^{ren}={V_\phi}{1\over 24}\left({2R\over
 \sqrt{\alpha^\prime}}+{{\sqrt{\alpha^\prime}}\over R}\right).}
The temperature dependent piece is given by the contribution of a
 two-dimensional massless scalar field, as expected from space-time
 considerations. 

\vskip .3in
\noindent
$\bullet\ ${\it Modding Type 0A by $G_2=\{1,\Omega(-1)^{F^L_s}\}$}
\vskip .2in
The action of $(-1)^{F^L_s}$ yields an extra minus sign in the RR
sector compared to \identifymodel . Therefore, for this model the
states invariant under the orientifold projection are $T$ and $C_+$. 

The partition function of the model is given by \partooa\ with the
replacement $\Omega\rightarrow \Omega(-1)^{F^L_s}$. Since $(-1)^{F^L_s}$
acts trivially in the NS-NS sector, the contribution from this sector
to the Klein bottle 
is the same as in the previous orientifold. Moreover, the RR sector
contribution to the Klein bottle also vanishes, due again to the
non-diagonal action of $\Omega(-1)^{F^L_s}$. Therefore, both the
crosscap state and the renormalized partition function 
are the same as in
the previous model and are given by  \factor\partioafinal. 

Despite many similarities, the fact that $C_+$ is invariant as opposed
to $C_-$ will result in important differences in constructing the
matrix model dual. 

\newsec{Unoriented ${\hat c}=1$ 0A Matrix Model}

The strategy here is to analyze the worldvolume theory on an unstable
D0-brane system. It is important to consider a brane configuration
that is unstable\foot{The theory also has uncharged D1-branes, but
they are stable, because the open string ``tachyon'' is actually massless.}, so that it has an open string  tachyon in its
spectrum, and we can recover the original closed string physics after
tachyon condensation.

The open string spectrum on a collection of D0-branes is stable in
Type 0A string theory. The D0-branes carry  charge under $C_-$, which
can be read off from the boundary state:
\eqn\bound{
|D0\ra={1\over \sqrt{2}}(|NSNS,-\ra+|RR,-\ra).}
The presence of the RR piece in the boundary state is responsible for
projecting
 out the open string tachyon
present in the bosonic string (and Type 0B string). This can be shown
by factorizing the cylinder amplitude into the annulus, which yields a
GSO projected open string partition function:
\eqn\openpart{
Z=\hbox{Tr}_{NS\oplus R}\left({1+(-1)^F\over 2}e^{-tH_o}\right).}
Therefore, we must consider a $\hbox{D0}-\overline{\hbox{D0}}$ system in order to
construct the dual matrix model. As in \DouglasUP , we will study a collection
of $N$ D0-branes and $N+m$ $\overline{\hbox{D0}}$-branes.
As usual, the GSO projection for
$\hbox{D0}-\overline{\hbox{D0}}$ open strings is reversed compared to that 
for D0-D0 open strings so that a tachyon
appears in 
the spectrum. The effective theory \TakayanagiSM\DouglasUP\ is described by a matrix quantum
mechanics with a $U(N)\times U(N+m)$ gauge field and a complex tachyon
in the bifundamental representation:
\eqn\spect{
A=\pmatrix{A_1&0\cr 0& A_2}\qquad M=\pmatrix{0&t\cr {t^\dagger}& 0}.}
After gauge-fixing, the problem reduces to studying the dynamics of $N$ non-interacting fermions moving in a plane
with angular momentum $m$  and subject to an inverted harmonic oscillator potential \DouglasUP. 

We now consider the orientifold models.

\vskip .3in
\noindent
$\bullet\ ${\it Modding Type 0A by $G_1=\{1,\Omega\}$}
\vskip .2in
Since $C_-$ is invariant under the action of $\Omega$, branes are mapped to
branes, and anti-branes to anti-branes,  under this symmetry. This means that we
can mode out by $G_1$ an arbitrary  $\hbox{D0}-\overline{\hbox{D0}}$
configuration (i.e. $m$ can be arbitrary). 

Demanding that the open string vertex operators are invariant under
the action of $g=\Omega$ results in the following
projections:
\eqn\projectopenstrings{\eqalign{
\gamma(g)A^t\gamma(g)^{-1}&=-A,\cr
\gamma(g)M^t\gamma(g)^{-1}&=M,}}
where $A$ and $M$ are given in \spect. There are two choices for the
representation matrices $\gamma(g)$, depending on the type of
orientifold. They are given by:
\eqn\choice{\eqalign{
O1^-:&\ \gamma(g)=\pmatrix{I&0\cr 0&I}\cr
O1^+:&\ \gamma(g)=\pmatrix{\pmatrix{0&iI\cr
-iI&0} &0\cr
0&\pmatrix{0&iI\cr-iI&0}}.}}
The solution to \projectopenstrings\ for the gauge field yields
\eqn\solgauge{\eqalign{
O1^-:&\ A\ \hbox{is in the}\ SO(N)\times SO(N+m)\ \hbox{Lie algebra}\cr
O1^+:&\ A\ \hbox{is in the}\ Sp(N)\times Sp(N+m)\ \hbox{Lie algebra},}}
while the solution for $M$ yields a field $t$ in the bifundamental
representation of the corresponding gauge group satisfying the following
reality conditions:
\eqn\soltach{\eqalign{
O1^-:&\ t=t^*, \cr
O1^+:&\ t=-Jt^*J,}}
where $J$ is the canonical $Sp(N)$ invariant tensor. 

\vskip .3in
\noindent
$\bullet\ ${\it Modding Type 0A by $G_2=\{1,\Omega (-1)^{F^L_s}\}$}
\vskip .2in
We already saw in the previous section that  $C_-$ is odd under the
action of $g=\Omega (-1)^{F^L_s} $. This means that a Type 0A D0-brane gets mapped to
a $\overline{\hbox{D0}}$-brane under the action of $g$. Therefore, in order
to be able to mod out Type 0A by $G_2$ in the presence of a
$\hbox{D0}-\overline{\hbox{D0}}$ system, we must demand that $m=0$, so that
there is an equal number of branes and anti-branes and the
configuration is $g$-symmetric. 

Since $g$ acts by mapping $00$-strings to ${\bar 0}{\bar
0}$-strings, the two factors in the $U(N)\times U(N)$ gauge group are identified,
and we are left with a $U(N)$ gauge field. We now
analyze the effect of the projection on $M$ \spect. There are two
inequivalent choices of how to project the open string spectrum,
depending on whether the representation matrix $\gamma(g)$ acting
on the Chan-Paton indices is symmetric or antisymmetric. The
resulting models are:
\eqn\modelosextra{\eqalign{
O1^-:&\ \gamma(g)=\pmatrix{0&I\cr I&0}\qquad
\gamma(g)M^t\gamma(g)^{-1}=M\ \Longrightarrow t\ \hbox{symmetric}\cr
O1^+:&\ \gamma(g)=\pmatrix{0&iI\cr -iI&0}\qquad
\gamma(g)M^t\gamma(g)^{-1}=M\ \Longrightarrow t\ \hbox{antisymmetric}.}}
By solving the projection we find that the matrix model dual to the
this orientifold model is a $U(N)$ gauge field together with a complex
symmetric or antisymmetric matrix $t$, corresponding respectively to
the model with an $O1^-$ or $O1^+$ orientifold. 

Next we perform the gauge-fixing procedure for these matrix models.
We start with the rectangular 
(or quiver) matrix models dual to the orientifold projections by
$G_1$. Using the action of the gauge
symmetry, one can bring the $N\times (N+m)$ rectangular 
matrix $M$ to an $N\times N$ diagonal
matrix with the rest of entries vanishing. One can then compute the
Jacobian for the change of coordinates. The result is \MorrisCQ\AndersonNW : 
\eqn\jacoboa{\eqalign{
U(N)\times U(N+m)&:\ J=\prod_i
\lambda_i^{1+2m}\prod_{k<l}(\lambda_k^2-\lambda_l^2)^2\cr
SO(N)\times SO(N+m)&:\ J=\prod_i
\lambda_i^{m}\prod_{k<l}(\lambda_k^2-\lambda_l^2)\cr
Sp(N)\times Sp(N+m)&:\ J=\prod_i
\lambda_i^{4m+3}\prod_{k<l}(\lambda_k^2-\lambda_l^2)^4.}}
The Schr\"odinger equation that needs to be solved is \hamil. By
redefining the wavefunction $\Psi(\lambda)=\sign(J) J^{-1/2}f(\lambda)$, we
find that $f(\lambda)$ is an eigenfunction of the following
Hamiltonian:

\eqn\Hamilparta{\eqalign{
{\tilde H}&=-{1\over 2}\sum_i {d^2\over
d\lambda_i^2}+{\alpha}({\alpha\over 2}-1)\sum_{i<j}\left({1\over
(\lambda_i-\lambda_j)^2}+{1\over (\lambda_i+\lambda_j)^2}
\right)\cr
&+{\alpha\over 2}(m+b)({\alpha\over 2}(m+b)-1)\sum_i{1\over
\lambda_i^2}+\sum_i
U(\lambda_i).}}
The parameters $\alpha$ and $b$ depend on the model and are displayed in Table 1.
This Hamiltonian describes the so-called $BC_N$ Calogero-Moser system. Its complete integrability
on the quantum level has been demonstrated in \CMBCone\CMBCtwo . It should be possible to 
determine its free energy in the double-scaling limit using the asymptotic Bethe ansatz; we leave this as a problem for
the future.

\input tables
\bigskip
\begintable
|~~~~$U$~~~~|~~~~$SO$~~~~|~~~~$Sp$~~~~\crthick
~~~$\alpha$~~~|$2$|$1$|$4$\crthick
~~~$b$~~~|$1/2$|$0$|$3/4$\endtable
\centerline{{\bf Table 1}:\ {\ninerm Values of parameters for the
various models.}}
\noindent

One can analyze the matrix model for the orientifold of Type 0A by
$G_2$ in a similar way. The gauge group is $U(N)$, and the tachyon field $M$
is
a symmetric or antisymmetric complex matrix. Reducing to the eigenvalues,
we find the following Jacobians:
\eqn\Jac{\eqalign{
M^t=M &:\ J=\prod_i \lambda_i \prod_{k<l} (\lambda_k^2-\lambda_l^2),\cr
M^t=-M &:\ J=\prod_i \lambda_i \prod_{k<l} (\lambda_k^2-\lambda_l^2)^4.}}
The corresponding Hamiltonian is given by Eq.~\Hamilparta\ with $m=0,$
$\alpha=1,$ $b=1$ (for $M^t=M$) and $m=0,$ $\alpha=4,$ $b=1/4$ (for
$M^t=-M$). In particular, the quantum mechanics of the eigenvalues is
integrable.

\bigbreak\bigskip\bigskip\centerline{{\bf Acknowledgements}}\nobreak
We would like to thank Juan Maldacena for useful
discussions.
J.~G.~was supported by the Sherman Fairchild Prize Fellowship. 
This research was supported in part by the DOE grant
DE-FG03-92-ER40701.

\vfill

\listrefs

\end